**KNIMEZoBot: Enhancing Literature Review with Zotero and KNIME OpenAI Integration using Retrieval-Augmented Generation.**

Suad Alshammari[1, 2], Lama Basalelah[1, 3], Walaa Abu Rukbah[1, 4], Ali Alsuhibani[1, 5] and Dayanjan S. Wijesinghe[1, 6, 7.]

**1**. Department of Pharmacotherapy and Outcomes Sciences, School of Pharmacy, Virginia Commonwealth University. **2**. Faculty of Pharmacy, Northern Border University, Saudi Arabia. **3**. Faculty of Pharmacy, Imam Abdulrahman Bin Faisal University, Saudi Arabia. **4**. Faculty of Pharmacy, University of Tabuk, Saudi Arabia. **5**. Department of Pharmacy Practice, Unaizah College of Pharmacy, Qassim University, Unaizah, Saudi Arabia. **6**. Institute for Structural Biology, Drug Discovery and Development, Virginia Commonwealth University, Richmond, Virginia, USA. **7**. Da Vinci Center, School of Pharmacy, Virginia Commonwealth University School of Medicine, Richmond, Virginia, USA.

Project files to be found at: https://github.com/dayanjan-lab/KNIMEZoBot

**Abstract:**

Academic researchers face challenges keeping up with exponentially growing published findings in their field. Performing comprehensive literature reviews to synthesize knowledge is time-consuming and labor-intensive using manual approaches. Recent advances in artificial intelligence provide promising solutions, yet many require coding expertise, limiting accessibility. KNIMEZoBot represents an innovative integration of Zotero, OpenAI, and the KNIME visual programming platform to automate literature review tasks for users with no coding experience. By leveraging KNIME's intuitive graphical interface, researchers can create workflows to search their Zotero libraries and utilize OpenAI models to extract key information without coding. Users simply provide API keys and configure settings through a user-friendly interface in a locally stored copy of the workflow. KNIMEZoBot then allows asking natural language questions via a chatbot and retrieves relevant passages from papers to generate synthesized answers. This system has significant potential to expedite literature reviews for researchers unfamiliar with coding by automating retrieval and analysis of publications in personal Zotero libraries. KNIMEZoBot demonstrates how thoughtfully designed AI tools can expand accessibility and accelerate knowledge building across diverse research domains.

**Introduction**:

The current era witnesses academicians, clinicians, and researchers grappling with the formidable challenge of information overload[1,2]. The incessant surge in published research findings over recent years has significantly outpaced the ability to stay updated. This challenge is poised to intensify with the advent of natural language optimized AI technologies, which are expected to propel the pace of discoveries and subsequent publications at an even faster rate[3]. The dire need for a mechanism to proficiently manage and assimilate this burgeoning knowledge is palpable. Within this complex quandary, two distinct scenarios emerge:

Firstly, the task of extracting precise answers from a pre-existing knowledge corpus poses a hurdle[1]. Researchers typically amass publications pertinent to their expertise in reference libraries. This textual corpus expands over time with the continual influx of new findings. The task of extracting specific information from this meticulously curated content escalates in complexity with the growing volume of publications, compelling users to sift through numerous documents. Consequently, the appeal for an AI-driven platform capable of synthesizing information from multiple different publications to precise queries from an ever-expanding corpus of curated scientific literature within personal and group reference libraries is burgeoning.

Secondly, the endeavor of encapsulating knowledge through exhaustive literature reviews unveils knowledge gaps and unveils avenues for consequential research[4]. This endeavor entails the conduct of meticulous literature reviews which begin with the identification and collation of relevant publications to address the posed inquiries. Post collection, a thorough analysis of the amassed information is essential to derive answers to specific queries. The conventional workflow of executing literature reviews is notably time-consuming and labor-intensive[5]. Given the swift pace of new discoveries, the traditional approach often yields a knowledge summary that becomes obsolete by the time of its completion.

The imperativeness for simplified, automated strategies enabling researchers to query curated literature libraries and routinely refresh their domain knowledge is apparent. The recent strides in artificial intelligence (AI), particularly the Large Language Models (LLMs), harbor the potential to alleviate the aforementioned challenges[6]. Platforms like ChatGPT, Claude, or Bard are proficient in summarizing papers and synthesizing findings across multiple documents[7]. However, their native "chat" formats present certain impediments for academic research. These include constraint of context window lengths[8] and the propensity for confabulation (hallucination)[9]. For instance, publicly available chatGPT4 has a context window of about 5000 words, while Claude's window extends to 75,000 words, rendering a bulk of academic publications too lengthy for chat GPT. Claude, although capable of summarizing single publications, finds its utility curtailed across multiple documents.

A notable breakthrough addressing the context window limitations and hallucination and aiding in data summarization from diverse documents is the Retrieval Augmented Generation (RAG) approach[10,11]. The operational workflow of RAG commences with the segmentation of broad text corpora into smaller, overlapping textual fragments. Following this, these fragments are transformed into vector representations and cataloged within a vector-based database. Upon query submission, it's converted into a vector form. A vectorial similarity assessment is then executed to identify all text segments within the database showcasing semantic alignment. The query, along with all relevant text fragments, is forwarded to a Large Language Model (LLM) to formulate a coherent and pertinent response. This technique adeptly navigates the context window constraints, fostering response synthesis across varied domains. The capability to repeatedly execute this process establishes a robust question-and-answer framework, invaluable for extensive literature reviews, serving scholars and medical professionals.

The RAG-based system, although expedient in summarizing information, hitherto necessitated substantial coding knowledge. Recognizing that a sizable faction of academicians and clinicians lacks coding expertise, we orchestrated a code-free, open-source strategy, culminating in the creation of KNIMEZoBot. This innovation amalgamates three pivotal elements: Konstanz Information Miner (KNIME) - a code-free data science platform, Zotero - an open-source reference management system, and GPT4 from Open AI - the chosen language model for knowledge synthesis. KNIMEZoBot heralds a revolutionary stride in enhancing the literature review workflow by seamlessly integrating the prowess of reference managers, scholarly databases, and AI. Through this ingenious approach, we have democratized access to AI-powered research tools, opening doors for those with non-coding backgrounds to harness natural language queries for interacting with the curated publications housed in their Zotero libraries, thereby significantly amplifying the accessibility and utility of AI in academic spheres.

## Materials and Methods

**KNIME:** For the development of the KNIMEZoBot platform, KNIME played an integral role in providing the graphical modular interface for building the workflow steps, integrating the Zotero and OpenAI APIs seamlessly via dedicated nodes, processing the extracted text data, creating the FAISS vector indexing workflow, and hosting the final chatbot user interface. KNIME is an open-source platform originating from the University of Konstanz in Germany, catering to data analytics, reporting, and integration needs, with a strong footing in data science and machine learning domains[12]. It's a free, community-enhanced tool, widely embraced by data professionals globally owing to its user-friendly, graphical interface enabling code-free workflow creation, modification, and visualization. KNIME’s core strength is its extensive node repository facilitating seamless data pipeline construction for tasks ranging from data preprocessing to advanced analytics using a non/low code approach. It boasts robust data integration, connecting effortlessly to various data sources like databases and web services, thus centralizing data for comprehensive analysis. Scalability is a hallmark of KNIME, adeptly managing small to large datasets, with ease of integration into big data frameworks like Apache Hadoop and Apache Spark. The platform supports building, training, and evaluating machine learning models utilizing popular libraries such as scikit-learn and TensorFlow, alongside offering an array of statistical and analytical techniques. Automation is seamless with KNIME, allowing scheduled workflow executions, while its server facilitates collaborative efforts and workflow sharing. Commercial versions of KNIME extend advanced features and support, enriching its open-source ecosystem. It's a versatile tool for creating insightful reports, visualizing data, and finds applications across diverse fields including bioinformatics, predictive analytics, business intelligence, and industrial research.

**KNIME extensions used:**

*KNIME AI Extension (Labs)*[13]: The KNIME labs extension enables users to leverage powerful large language models (LLMs) from OpenAI, Hugging Face Hub, and GPT4ALL for tasks like chat and text embeddings. It also provides connectivity to vector stores like Chroma and FAISS for building knowledge bases that can inform chatbot responses. The extension allows combining vector stores and LLMs into intelligent agents. These agents can dynamically select the most relevant vector store to query based on the user input, enabling more natural and knowledgeable conversations. Overall, this extension brings together state-of-the-art LLMs and vector stores within the KNIME analytics platform, unlocking new possibilities for building conversational interfaces and knowledge-powered AI assistants.

*KNIME REST Client Extension*[14]: The KNIME REST Client Extension provides nodes for making REST API calls within KNIME workflows. This enables seamless integration with web services and APIs.

The Get Request node[15] is used to send HTTP GET requests to REST endpoints. It allows specifying the URL, headers, query parameters, and authentication settings. The response from the REST API is returned as a JSON/XML document that can be further processed in the KNIME workflow[14].

*KNIME Python 2 Integration (legacy)*[16]: This extension encompasses the legacy version of KNIME Python integration. It facilitates the integration of Python 2 and Python 3 scripts within the KNIME platform. The extension operates by executing Python scripts in a local Python installation, which is not included in the extension package.

*KNIME Python Integration*[17] : The "KNIME Python Integration" is the modern and preferred choice for Python integration within KNIME. This extension incorporates nodes that enable the execution of Python 3 scripts seamlessly in the KNIME workflow. Notably, this integration brings substantial performance improvements compared to its legacy counterpart. It also provides enhanced support for handling larger-than-memory datasets. Additionally, this extension comes equipped with a Python installation that includes a curated selection of essential Python packages.

Note: Throughout our workflow, we employed both the "KNIME Python 2 Integration (legacy)" and the "KNIME Python Integration" extensions interchangeably.

*KNIME JSON-Processing*[18]: The KNIME JSON-Processing extension provides nodes for working with JSON data within KNIME workflows. It allows for parsing, creating, transforming, and serializing JSON documents. The JSON To Table node enables easy ingestion of JSON data into tabular form for use in KNIME workflows. It reduces the complexity of handling nested JSON structures and schemas.

*Python (Version 3.9)*[19]: Integrating of Python and Anaconda with KNIME can be a powerful combination that allows data scientists and analysts to leverage the extensive libraries and capabilities of Python within the KNIME analytics platform. This integration provides a seamless way to utilize Python scripts, packages, and machine learning models within the KNIME workflows. A detailed guide for Python integration in KNIME has been published elsewhere[20]

Specifically, Python nodes were utilized to execute API calls to extract metadata and PDF files from the Zotero reference manager using its REST API bindings. The Python Requests library facilitated sending GET requests to the API and processing the responses. Another vital usage was the Langchain library[21] within a Python node to load in the full text of PDF papers and segment them into smaller chunks that meet the length limits of the GPT model inputs.

The Python nodes accept inputs from earlier workflow components, run the defined Python logic and code using those inputs, and return any outputs to subsequent nodes in the workflow. For instance, a node might accept a list of extracted PDFs from the Zotero API calls, utilize Langchain to chunk each PDF into shorter text segments, and output these chunks to the next node for vectorization.

The following Python packages were installed and imported within the Python nodes in KNIME to support core functionality. The installation can be done in multiple ways; we used the following:

1- Open Anaconda Prompt from the Start menu.
2- Write this command: "conda activate <your_environment>." Your_environment is the environment name that is set up in the KNIME Python preference.
3- After the name is changed from base to the name of the environment, install the following libraries via pip:
    - pip install pandas openai langchain unstructured fitz PyPDF2 PyMuPDF "unstructured[pdf]"

**Zotero:**

Zotero, a free, open-source reference management software[22], is cherished by a broad spectrum of academia and professionals for easing the collection, organization, and citation of research materials. Originating from George Mason University, it’s a boon for scholarly research and writing, streamlining reference, citation, and bibliography management. Key facets include effortless reference collection from diverse sources like websites and academic journals, with automatic citation information extraction from web pages and PDFs. Its intuitive interface facilitates organizing references via folders, tags, and notes, ensuring easy retrieval. A hallmark feature is its citation and bibliography generation in numerous styles like APA and MLA, significantly reducing formatting time. Integration with prevalent word processors like Microsoft Word and Google Docs allows direct citation insertion and bibliography generation in documents, ensuring accuracy and consistency. Its PDF management capability lets users attach, organize, and annotate PDFs within the reference library. Zotero encourages collaborative research through shared library features, vital for research teams. It offers cloud synchronization for easy access across devices and data backup, enhancing data security. Browser extensions for Chrome and Firefox simplify capturing references online. Being open-source, it's continually evolved by community contributions, and its cross-platform availability extends its reach. Applications are vast, aiding academic research, education, library assistance, and professionals across legal, medical, and media fields in managing and citing a vast array of references effortlessly.

## Results and Discussion

### KNIMEZoBot:

The developed application "KNIMEZoBot", represents an innovative integration of Zotero and OpenAI through the code free platform KNIME to streamline literature reviews and research. This project seamlessly combines the above-mentioned Zotero reference manager, with OpenAI's powerful natural language processing capabilities via a RAG based approach using KNIME as the interface. The primary goal is to simplify retrieving PDFs from Zotero libraries and collections and then utilize OpenAI within KNIME workflows to ask insightful questions and extract key information from academic papers.

KNIMEZoBot uses a Retrieval-Augmented Generation (RAG) architecture, first conducting a semantic search to identify relevant passages from retrieved PDFs. It then leverages large language models (in this case OpenAI's GPT models) to synthesize natural language answers based on the extracted information. This enables KNIMEZoBot to provide informative responses to questions by efficiently searching academic papers and distilling salient facts and main points. Overall, the integration of Zotero and OpenAI represents an innovative approach to enhance literature reviews and research by combining reference management, scholarly databases, and OpenAI.

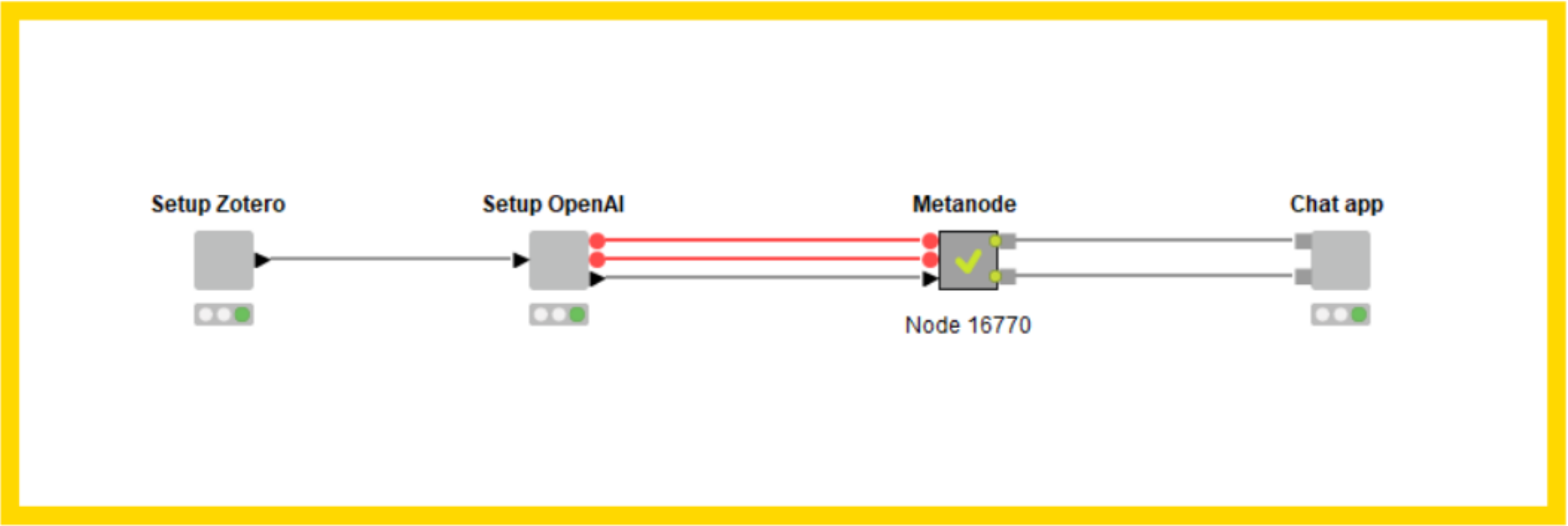

**Figure1:** Underlying overall workflow for KNIMEZoBot.

### First component (Setup and configure Zotero):

In order to effectively use the KNIMEZoBot, users need to follow a series of key steps. The first requirement is selecting the type of Zotero library they want to access - either a personal

Zotero library or a group library. Based on that choice, users will need to input their corresponding Zotero API key, which allows the system to interface with the library.

Additionally, users will need to provide either their personal Zotero user ID if accessing their own library, or the group ID if accessing a shared group library. To assist users in easily finding and copying their user ID or group ID, we have included hyperlinks within the system interface that direct users to Zotero guides with instructions on locating that information.

Furthermore, to enable more targeted searches, users have the option to filter based on Zotero collections. This allows them to refine the content being retrieved from their library down to specific collections, rather than everything in the library. The system was designed to be flexible - some users may want to search across their entire library, while others may want to narrow in on papers from select collections.

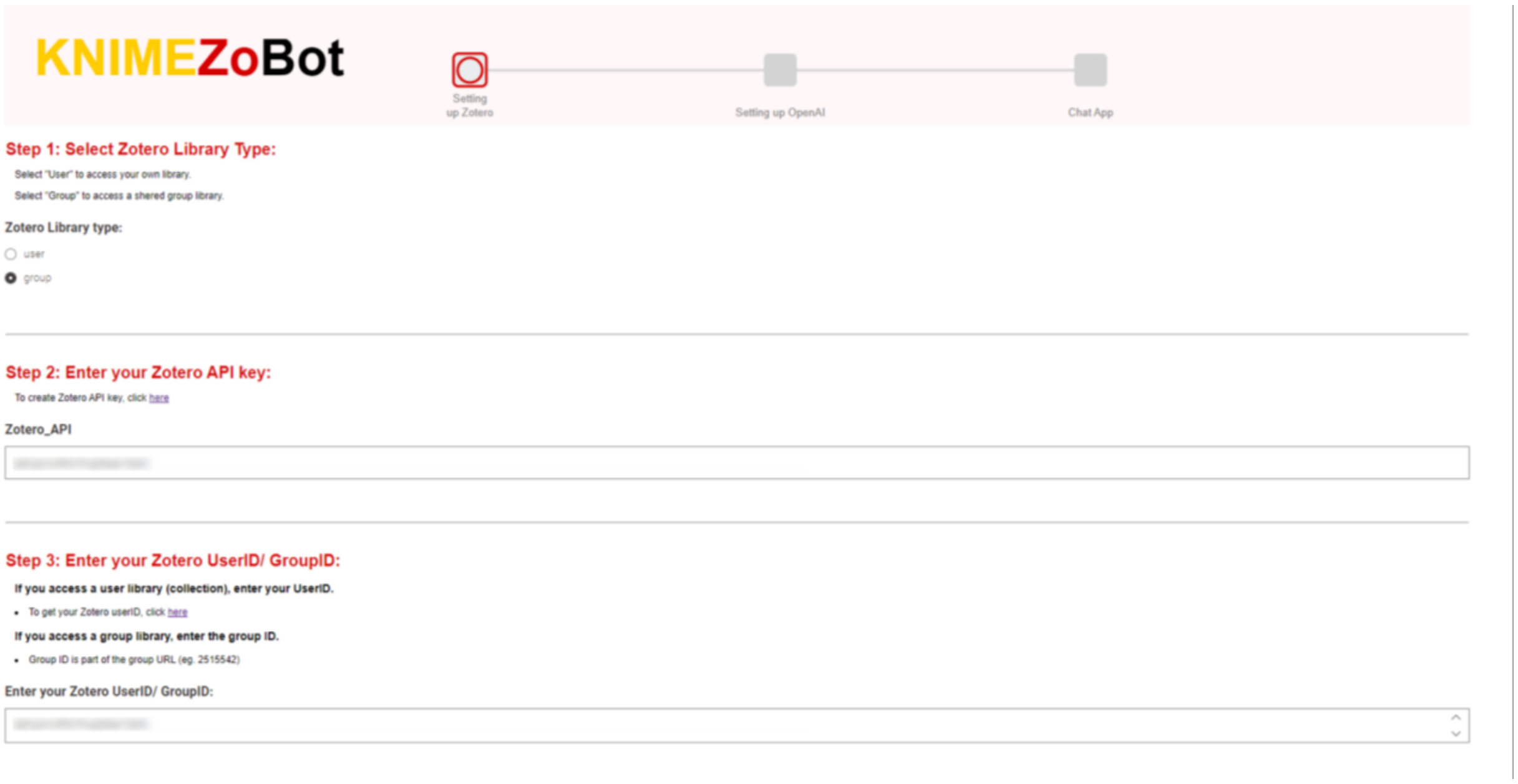

**Figure 2:** Component “Setup Zotero” when executed - User interface of KNIMEZoBot. Users are required to complete the Zotero information fields.

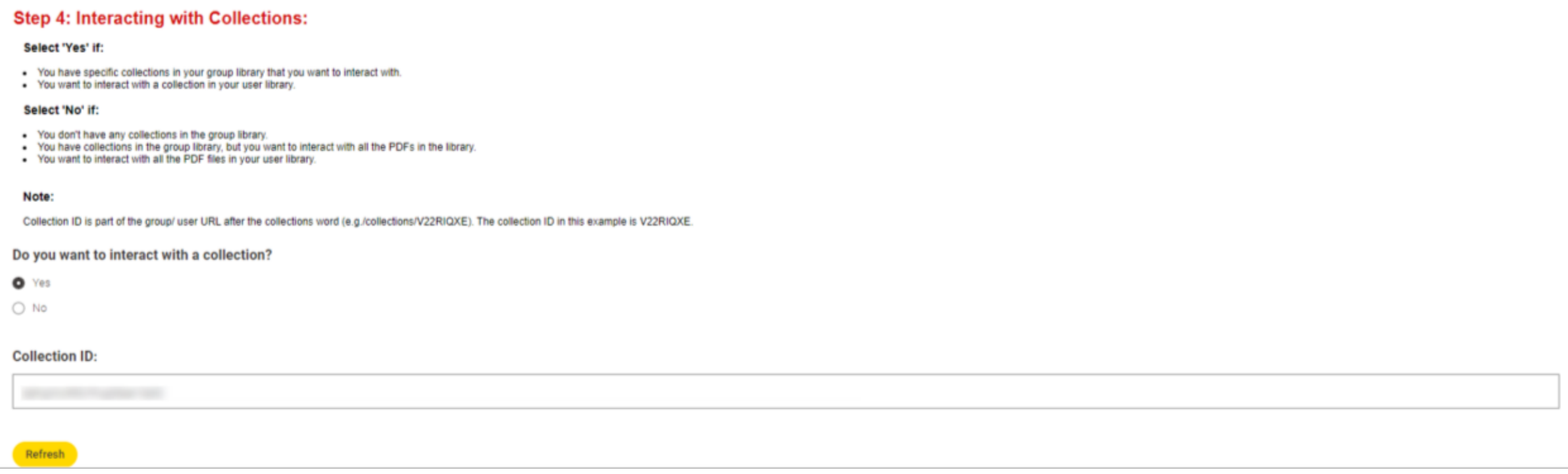

**Figure 3:** Continuation of the first component when executed. We provided options to filter by specific collections.

https://github.com/dayanjan-lab/KNIMEZoBot

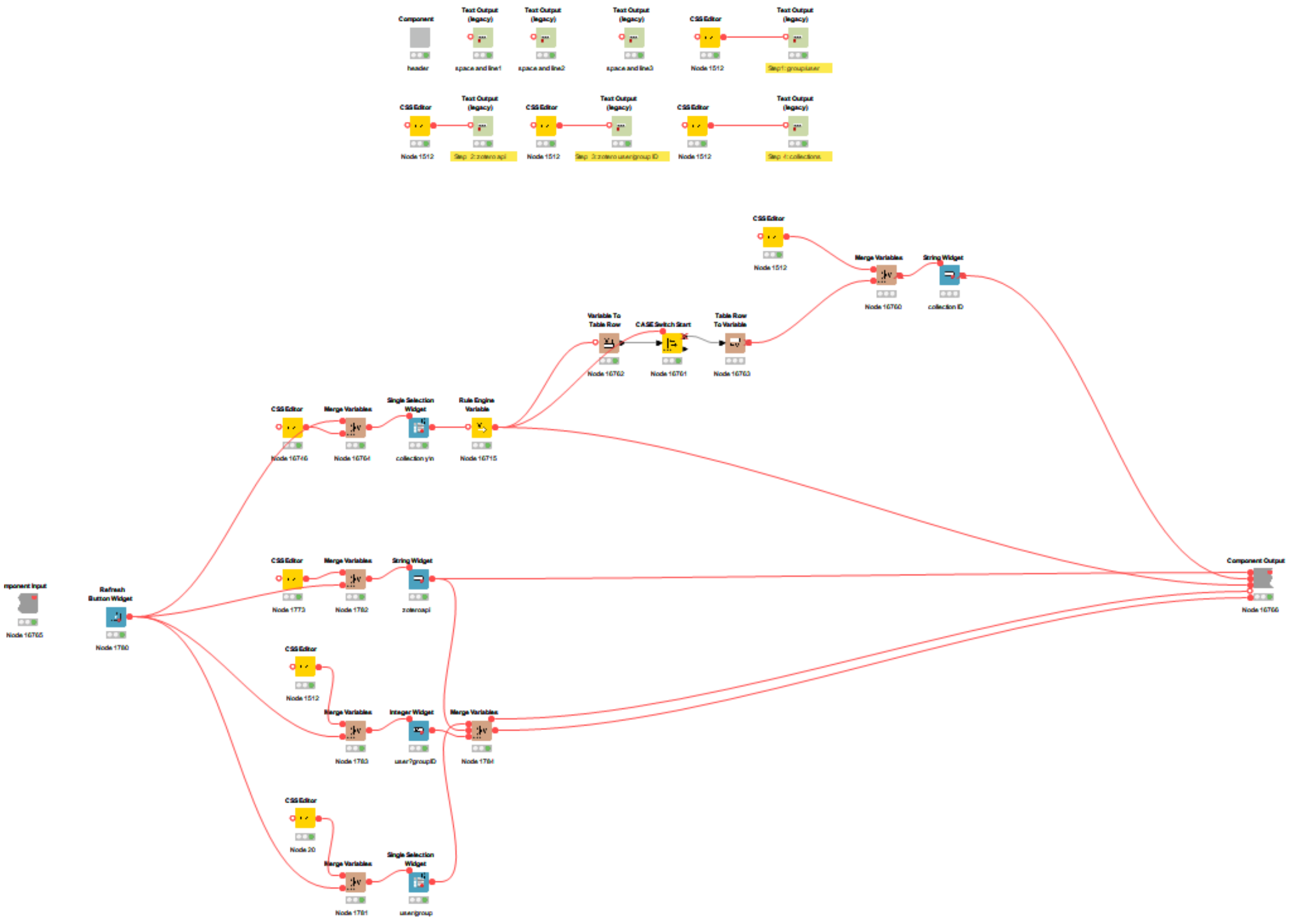

**Figure 4:** The "setup Zotero" component contains widget nodes[23] that allow the user to input the information required by the system. It also includes text output nodes and a header node that control the appearance of the user interface.

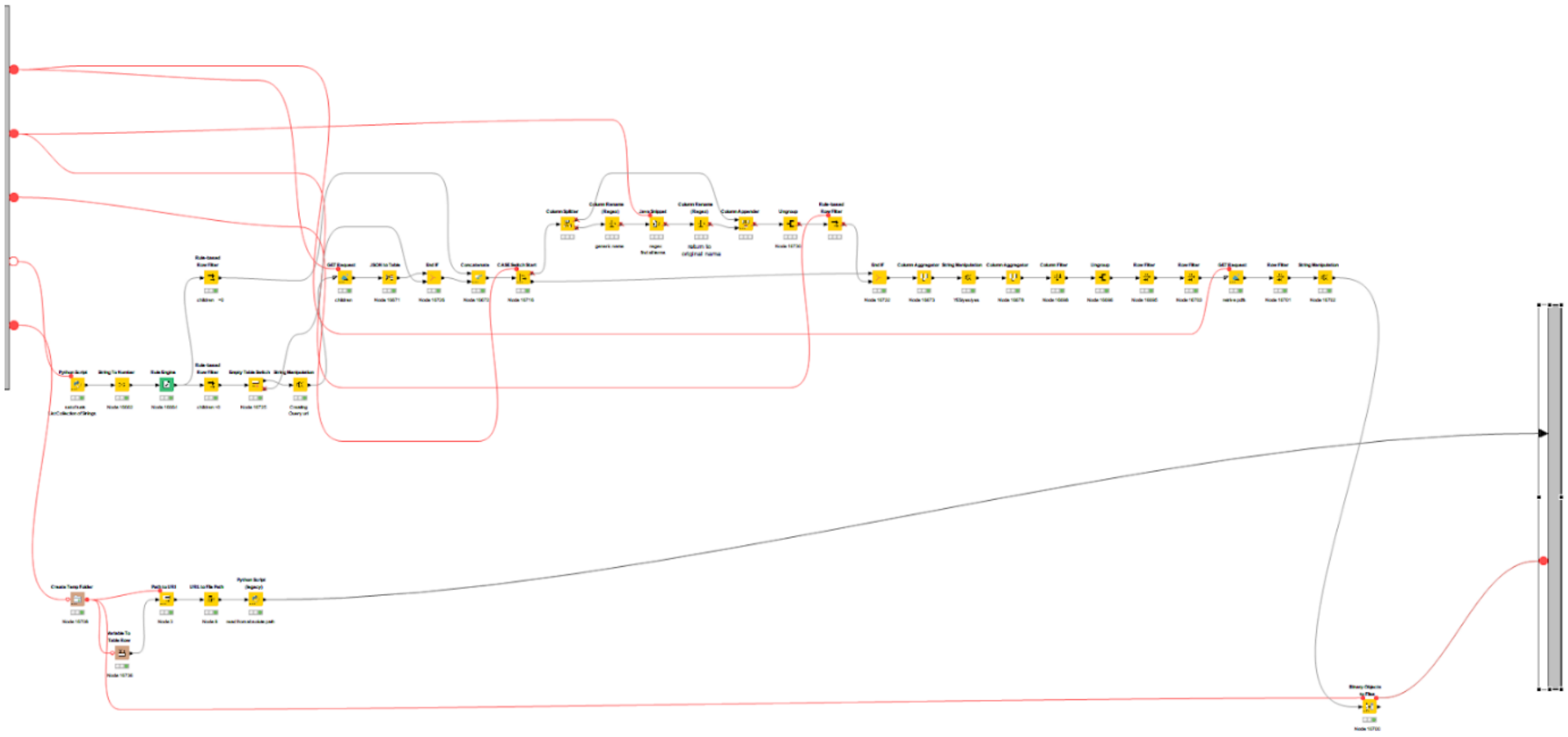

**Figure 5:** The first metanode processes information from the "setup zotero" component. Specifically, the “python script” node[24] accesses and extracts all data from the Zotero library. The “get request” node[15] (a child node) retrieves attached files for each Zotero item. Subsequently, the 'binary objects to files' node[25] is employed to facilitate the secure storage of PDF documents in a pre-defined temporary directory within the workflow.

**Second component (Setup OpenAI):**

The second core component of the system involves setting up the OpenAI environment according to the user's preferences. Users have the ability to adjust key settings such as chunk size and chunk overlap. Chunk size refers to the maximum number of tokens processed per API request, while overlap determines the number of duplicated tokens between chunks. Giving users control over these parameters enables them to customize the configuration based on their specific computational needs and use case.

After inputting their OpenAI API key, which grants access to the AI models, users can select from a variety of available models offered through the OpenAI API. Users can make a selection from a range of available OpenAI models, including but not limited to GPT-3.5 Turbo and GPT-4.

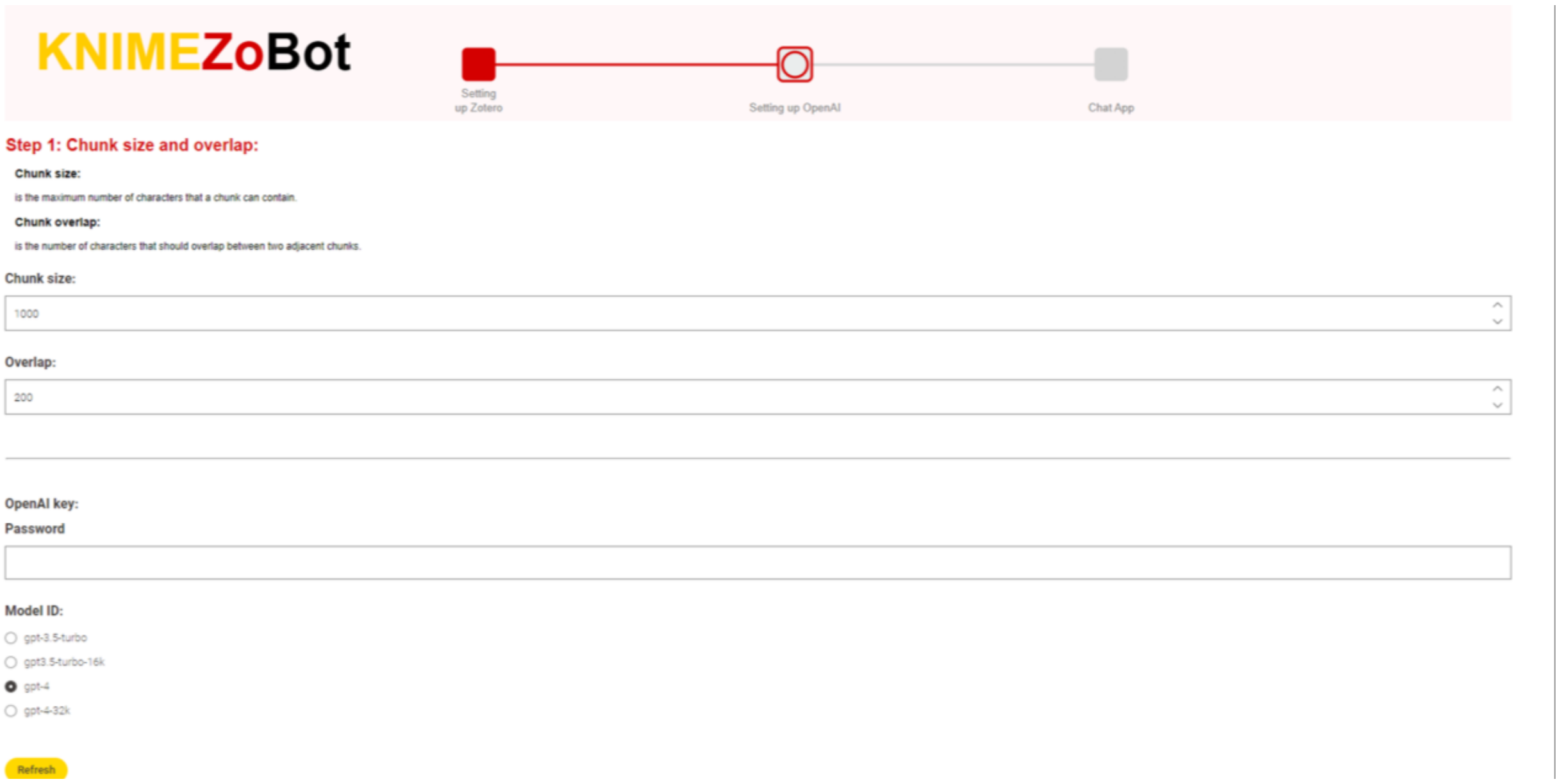

**Figure 6:** Component "Setup OpenAI" when executed- Users are required to select chunk size and overlap settings for text processing, enter their OpenAI API key, and select an AI model.

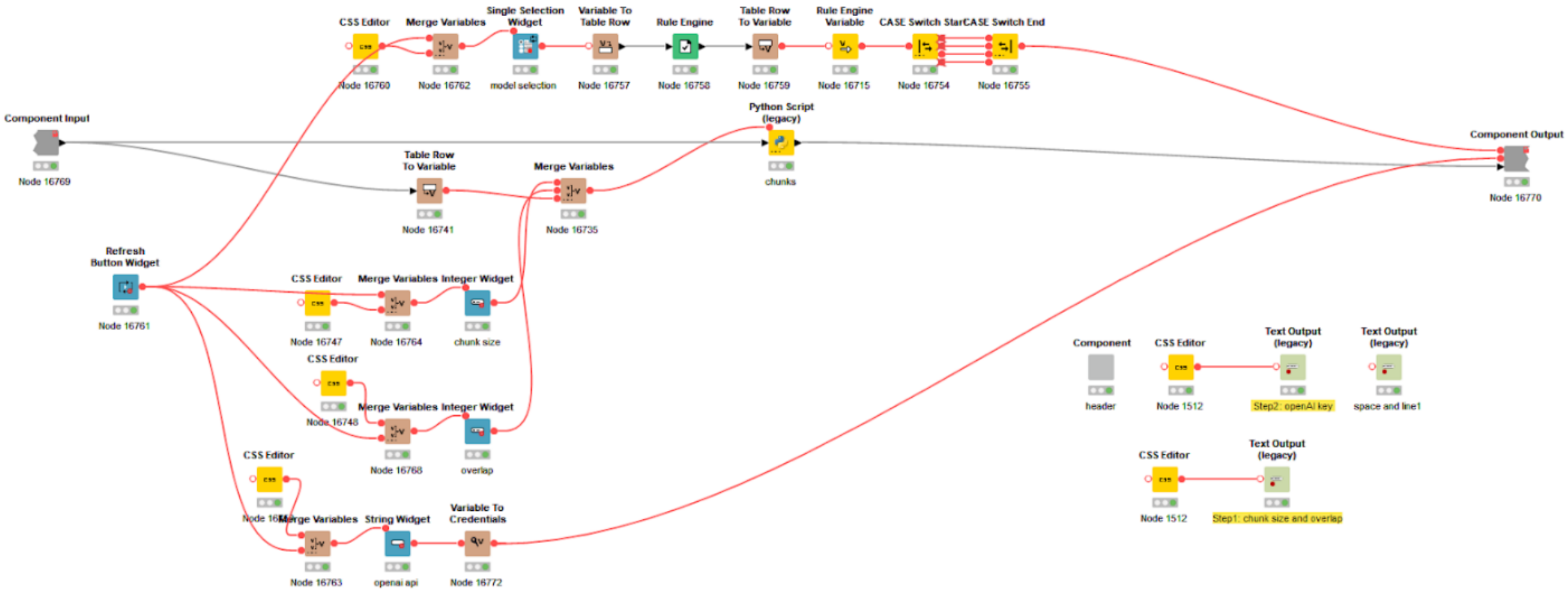

**Figure 7:** The "setup OpenAI" component contains widget nodes enabling users to input their OpenAI API key, chunk size, chunk overlap, and select an AI model as required by the system. Additionally, text output nodes and a header node are included for the appearance of the user interface. The "Python Script" node was used to read and split the PDFs to smaller chunks. We used the Langchain package[21] "unstructuredPDFLoader" to load the documents. The documents were then split into smaller chunks because of size limitation as GPT models have a maximum input size, usually 1024-2048 tokens. Breaking PDFs into smaller chunks allows to feed longer documents into the models.

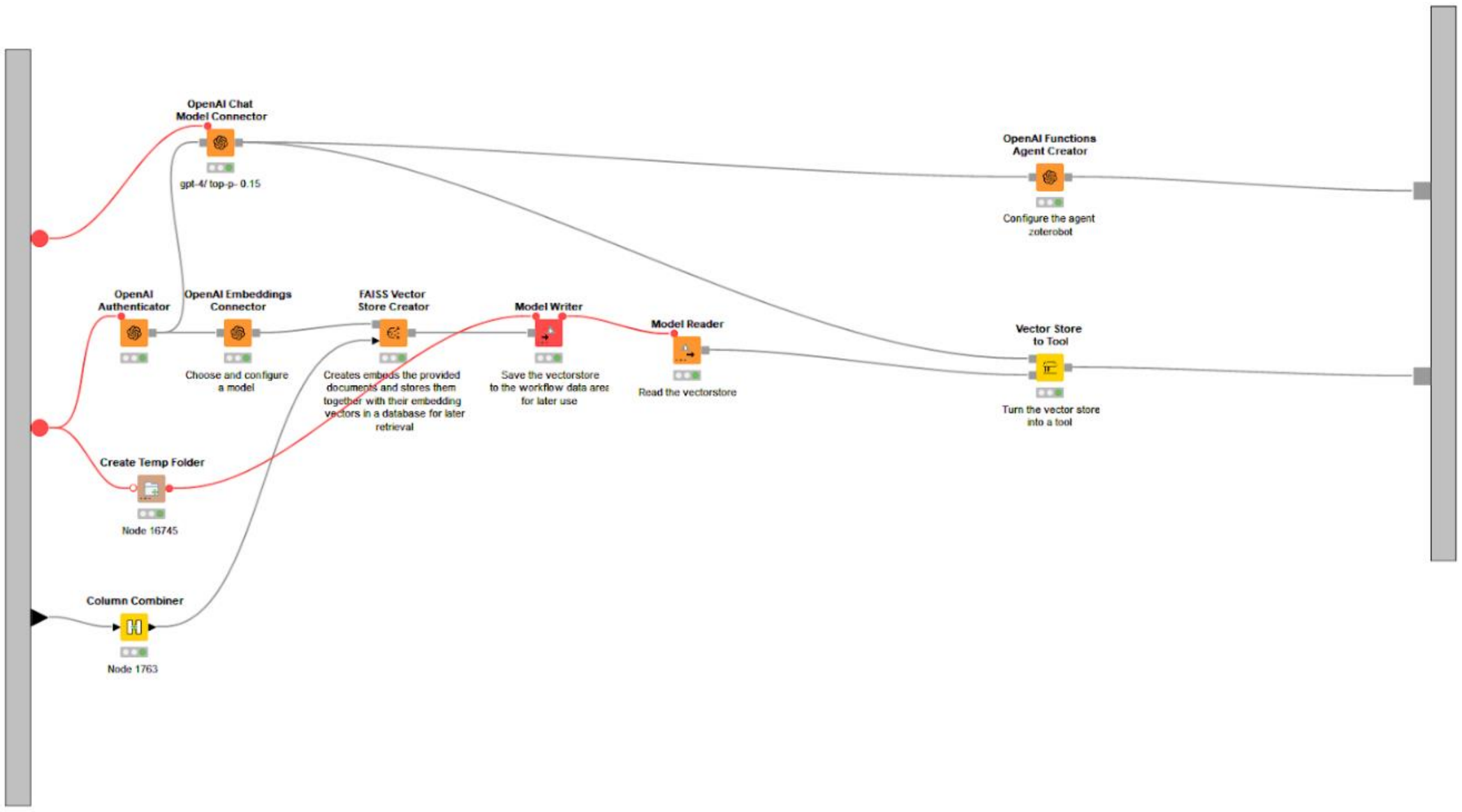

**Figure 8:** In the metanode, we used the "FAISS Vector Store Creator" node[26]. This node will store the numerical vector representation created by the embedding model from the "OpenAI Embeddings Connector" node[27]. Also, we selected "text-embedding-ada-002" as the embedding model.

The "OpenAI Functions Agent Creator" node[28] is used to customize the system message. We rote the following message "*You are KnimeZoBot, an AI assistant specifically designed to seamlessly integrate the power of the KNIME platform with the vast knowledge stored within your Zotero library. Your mission is to provide the user with a unique and efficient way to access information, answer questions, and streamline users' research tasks by tapping into your personal Zotero library. Get the answer only from the provided information and if it is not store there write "I apologize, but I do not have any information about it in my Zotero library*."

**Last component (Chat app):**

The last component of the system is the Chat application, which provides an interactive interface for users to engage with their Zotero library. This chatbot-style app enables users to pose questions and queries about the content of their Zotero library in a natural conversational format. The seamless integration of the chatbot with the Zotero reference database creates a convenient and user-friendly method for users to search for information within their library.

In addition, users have the option to download their full conversation history with the chatbot in a .csv format. This allows users to save all of their questions and the chatbot's responses so they can refer back to the information later.

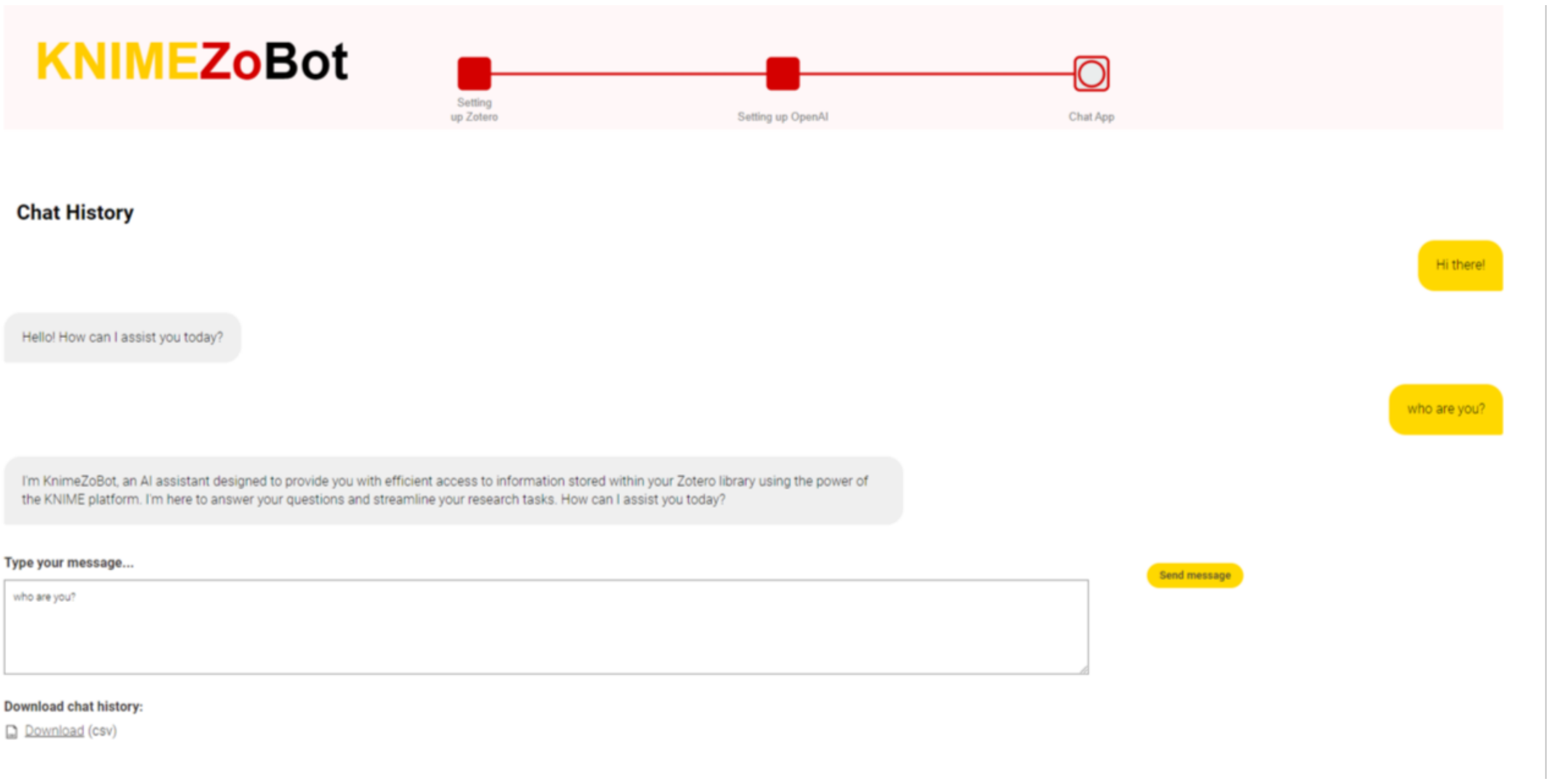

**Figure 9:** Final Component "Chat App"- Chat Interface when deployed. This component allows users to ask questions and receive answers through a conversational chatbot. Users can also download their full chat history as a CSV file.

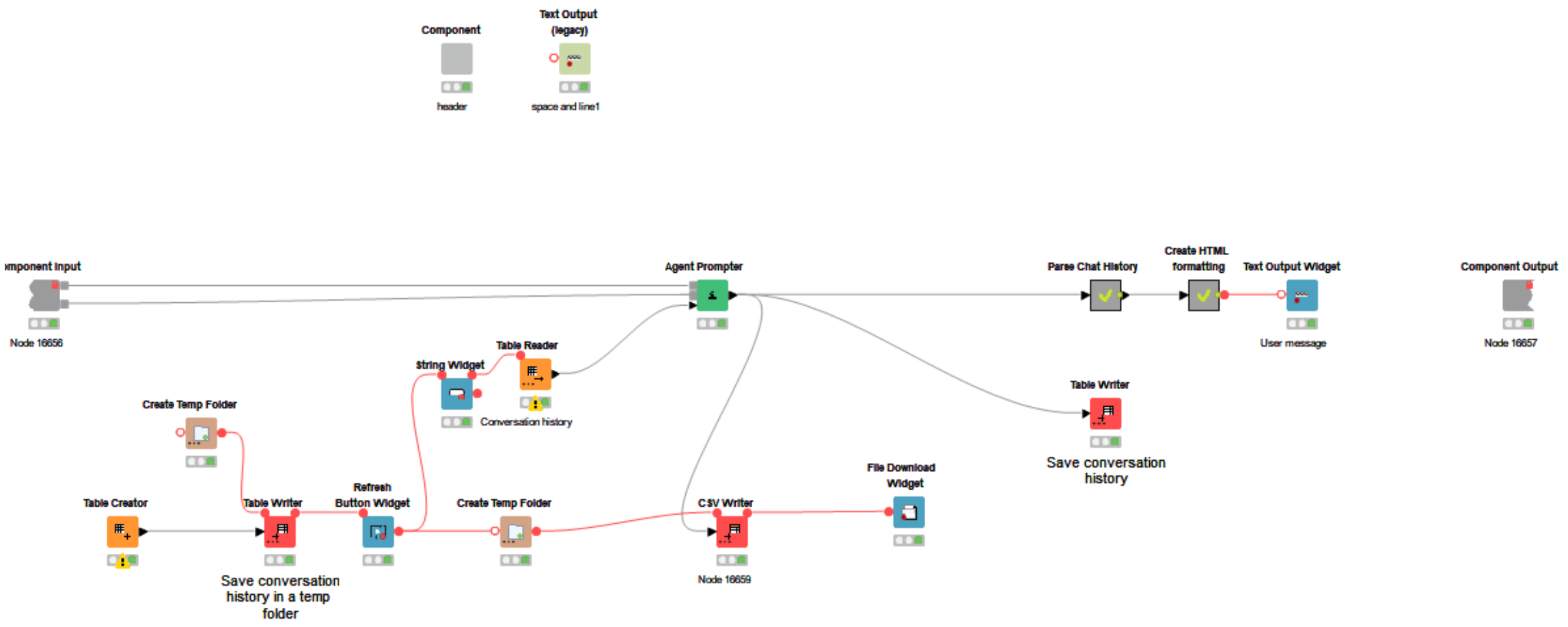

**Figure 10:** The final "Chat app" component utilizes an "Agent Prompter" node[29] to leverage an AI agent, prompts, and the conversation history from the input table to generate responses. The conversation table requires at least two string columns to store previous exchanges. Additionally, an option is provided to save the chat history.

**Conclusion:**

In summary, the KNIMEZoBot represents a promising integration of technologies to expedite literature reviews or undertake natural language queries of existing Zotero libraries. By unifying the capabilities of Zotero, OpenAI, and KNIME, this system automates laborious tasks such as combing through academic papers to identify relevant information. Researchers can save significant time while benefiting from state-of-the-art AI techniques for synthesizing knowledge in a low code manner. This innovation demonstrates the potential for AI to assume a greater role in accelerating informed research. While further enhancements to the accuracy and sophistication of the automated analysis remain desirable, KNIMEZoBot marks an important step toward streamlining access to critical information in existing literature by domain experts who are not coders by training. By facilitating more rapid and comprehensive understanding of prior work, this system could substantially benefit the research community and knowledge-building process.

https://github.com/dayanjan-lab/KNIMEZoBot

**References:**

1. Arnold M, Goldschmitt M, Rigotti T. Dealing with information overload: a comprehensive review. *Front Psychol*. 2023;14:1122200. doi:10.3389/fpsyg.2023.1122200

2. Bornmann L, Haunschild R, Mutz R. Growth rates of modern science: a latent piecewise growth curve approach to model publication numbers from established and new literature databases. *Humanit Soc Sci Commun*. 2021;8(1):1-15. doi:10.1057/s41599-021-00903-w

3. Kousha K, Thelwall M. Artificial intelligence to support publishing and peer review: A summary and review. *Learn Publ*. n/a(n/a). doi:10.1002/leap.1570

4. Paré G, Kitsiou S. Chapter 9 Methods for Literature Reviews. In: *Handbook of eHealth Evaluation: An Evidence-Based Approach [Internet]*. University of Victoria; 2017. Accessed November 3, 2023. https://www.ncbi.nlm.nih.gov/books/NBK481583/

5. Tay A. How to write a superb literature review. *Nature*. Published online December 4, 2020. doi:10.1038/d41586-020-03422-x

6. Wagner G, Lukyanenko R, Paré G. Artificial intelligence and the conduct of literature reviews. *J Inf Technol*. 2022;37(2):209-226. doi:10.1177/02683962211048201

7. Nashwan AJ, Jaradat JH. Streamlining Systematic Reviews: Harnessing Large Language Models for Quality Assessment and Risk-of-Bias Evaluation. *Cureus*. 15(8):e43023. doi:10.7759/cureus.43023

8. Stern J. GPT-4 Has the Memory of a Goldfish. The Atlantic. Published March 17, 2023. Accessed November 3, 2023. https://www.theatlantic.com/technology/archive/2023/03/gpt-4-has-memory-context-window/673426/

9. Sharun K, Banu SA, Pawde AM, et al. ChatGPT and artificial hallucinations in stem cell research: assessing the accuracy of generated references - a preliminary study. *Ann Med Surg 2012*. 2023;85(10):5275-5278. doi:10.1097/MS9.0000000000001228

10. Lewis P, Perez E, Piktus A, et al. Retrieval-Augmented Generation for Knowledge-Intensive NLP Tasks. Published online April 12, 2021. doi:10.48550/arXiv.2005.11401

11. Chen J, Lin H, Han X, Sun L. Benchmarking Large Language Models in Retrieval-Augmented Generation. Published online September 4, 2023. doi:10.48550/arXiv.2309.01431

12. Berthold MR, Cebron N, Dill F, et al. KNIME - the Konstanz information miner: version 2.0 and beyond. *ACM SIGKDD Explor Newsl*. 2009;11(1):26-31. doi:10.1145/1656274.1656280

13. KNIME AI Extension (Labs). KNIME Community Hub. Accessed November 7, 2023. https://hub.knime.com/knime/extensions/org.knime.python.features.llm/latest

14. KNIME REST Client Extension. KNIME Community Hub. Accessed November 7, 2023. https://hub.knime.com/knime/extensions/org.knime.features.rest/latest

15. GET Request. KNIME Community Hub. Accessed November 7, 2023. https://hub.knime.com/knime/extensions/org.knime.features.rest/latest/org.knime.rest.nodes.get.RestGetNodeFactory

16. KNIME Python 2 Integration (legacy). KNIME Community Hub. Accessed November 7, 2023. https://hub.knime.com/knime/extensions/org.knime.features.python2/latest

17. KNIME Python Integration. KNIME Community Hub. Accessed November 7, 2023. https://hub.knime.com/knime/extensions/org.knime.features.python3.scripting/latest

18. KNIME JSON-Processing. KNIME Community Hub. Accessed November 7, 2023. https://hub.knime.com/knime/extensions/org.knime.features.json/latest

19. Python Release Python 3.9.0. Python.org. Accessed November 7, 2023. https://www.python.org/downloads/release/python-390/

20. KNIME Python Integration Guide. Accessed November 7, 2023. https://docs.knime.com/2021-12/python_installation_guide/index.html#_introduction

21. Introduction | 🦜🔗 Langchain. Accessed November 7, 2023. https://python.langchain.com/docs/get_started/introduction

22. credits_and_acknowledgments [Zotero Documentation]. Accessed November 3, 2023. https://www.zotero.org/support/credits_and_acknowledgments#about_zotero

23. Explore the Wonderful World of KNIME Widgets. KNIME. Accessed November 7, 2023. https://www.knime.com/blog/mini-guide-widget-examples

24. Python Script. KNIME Community Hub. Accessed November 7, 2023. https://hub.knime.com/knime/extensions/org.knime.features.python3.scripting/latest/org.knime.python3.scripting.nodes.script.PythonScriptNodeFactory

25. Binary Objects to Files. KNIME Community Hub. Accessed November 7, 2023. https://hub.knime.com/knime/extensions/org.knime.features.base.filehandling/latest/org.knime.base.filehandling.binaryobjects.writer.BinaryObjectsToFilesNodeFactory

26. FAISS Vector Store Creator. KNIME Community Hub. Accessed November 7, 2023. https://hub.knime.com/knime/extensions/org.knime.python.features.llm/latest/org.knime.python3.nodes.extension.ExtensionNodeSetFactory$DynamicExtensionNodeFactory:e1168c28

27. OpenAI Embeddings Connector. KNIME Community Hub. Accessed November 7, 2023. https://hub.knime.com/knime/extensions/org.knime.python.features.llm/latest/org.knime.python3.nodes.extension.ExtensionNodeSetFactory$DynamicExtensionNodeFactory:3a4ffd4b

28. OpenAI Functions Agent Creator. KNIME Community Hub. Accessed November 7, 2023. https://hub.knime.com/knime/extensions/org.knime.python.features.llm/latest/org.knime.python3.nodes.extension.ExtensionNodeSetFactory$DynamicExtensionNodeFactory:232d61e6

29. Agent Prompter. KNIME Community Hub. Accessed November 7, 2023. https://hub.knime.com/knime/extensions/org.knime.python.features.llm/latest/org.knime.python3.nodes.extension.ExtensionNodeSetFactory$DynamicExtensionNodeFactory:378eea

https://github.com/dayanjan-lab/KNIMEZoBot